\documentclass[a4paper, 10pt, conference]{IEEEtran} %abooud
\DeclareMathAlphabet\mathbfcal{OMS}{cmsy}{b}{n} 
\usepackage{cite}
\usepackage[pdftex]{graphicx}
\usepackage{dsfont}
\usepackage{comment} % for block commenting
\usepackage[usenames,dvipsnames]{color}
\usepackage{acronym}
\usepackage{multirow}
\usepackage[caption=false]{subfig}
\usepackage{nth}
\usepackage[geometry]{ifsym}
\usepackage{amsfonts}
\usepackage{amsmath,amssymb}
\usepackage{wasysym} % to be able to insert crescent or leftmoon
\usepackage{mdwmath}
\usepackage{mdwtab}
\usepackage{balance}
\usepackage[hyphens]{url}
\usepackage{xcolor, soul}
\sethlcolor{SkyBlue}
\usepackage{placeins}
\usepackage{float}
\usepackage{amsmath}
\DeclareMathOperator*{\argmax}{arg\,max}
\DeclareMathOperator*{\argmin}{arg\,min}
\ifCLASSINFOpdf
\else
\fi 
\begin{document}
\def\user{i}  
\def\userOth{j}
\def\player{\mathcal{I}}
\def\numPlayer{N}
\def\strategy{S}
\def\strategyI{s_{\user}}
\def\strategyJ{s_{\userOth}}
\def\payoff{U}
\def\reelnum{\mathds{R}}
\def\game{\mathcal{G}}
\def\bestresp{\mathcal{B}}
\def\noise{\omega(t)}
\def\totalpower{\mathcal{P}}
\def\mypower{\psi}
\def\pu{\mathcal{M}}
\def\su{\mathcal{F}}
\def\rx{\mathcal{P}}
\def\supower{\psi}
\def\antennagain{\rho}
\def\infobit{{L}}
\def\datarate{{R}}
\def\infot{{M}}
\def\sinr{\gamma_{\user}}
\def\stateInd{H}
\def\subcspace{\Delta f}
\def\landa{\h}
\def\in{∈}
\def\offRate{\varphi}
\def\rollOff{\beta}
\def\potentialG{V}
\def\utility{\payoff(\offRate,\rollOff)}
\def\equSinr{\frac{p_{\user}}{p_j+\noise_0}}
\def\equDer{\frac{p_{\user}}{(p_j+\noise_0)^2}}
\newtheorem{definition}{Definition}
\def\figurePath{pics/}
\graphicspath{ {pics/} }
\def\figWidth{3.45}
\makeatletter
\newcommand*{\rom}[1]{\expandafter\@slowromancap\romannumeral #1@}
\makeatother
\newcommand*{\Scale}[2][4]{\scalebox{#1}{$#2$}}%

% Keywords command
\providecommand{\keywords}[1]
{
	\small	
	\textbf{\textit{Keywords:}} #1
}

\title{Relay Selection for 5G New Radio Via Artificial Neural Networks
}

\author{\IEEEauthorblockN{Saud Aldossari , Kwang-Cheng Chen }\\
\IEEEauthorblockA{Department of Electrical Engineering \\ University of South Florida Tampa, Florida 33620, USA \\ Email:  saldossari@mail.usf.edu, kwangcheng@usf.edu}}
\maketitle

\begin{abstract}
Millimeter-wave supplies an alternative frequency band of wide bandwidth to better realize pillar technologies of enhanced mobile broadband (eMBB) and ultra-reliable and low-latency communication (uRLLC) for 5G - new radio (5G-NR). When using mmWave frequency band, relay stations to assist the coverage of base stations in radio access network (RAN) emerge as an attractive technique. However, relay selection to result in the strongest link becomes the critical technology to facilitate RAN using mmWave.
An alternative approach toward relay selection is to take advantage of existing operating data and apply appropriate artificial neural networks (ANN) and deep learning algorithms to alleviate severe fading in mmWave band. In this paper, we apply classification techniques using ANN with multilayer perception to predict the path loss of multiple transmitted links and base on a certain loss level, and thus execute effective relay selection, which also recommends the handover to an appropriate path. ANN with multilayer perception are compared with other ML algorithms to demonstrate effectiveness for relay selection in 5G-NR. 

\end{abstract}

\keywords{Machine Learning, Wireless Communications, MmWave, Neural Network, Multilayer Perceptrons, Classification, Relay Selection, SVM and Logistic Regression, 5G-NR.}

\acrodef{icic}   [ICIC]   {inter-cell interference coordination}
\acrodef{ne}     [NE]     {Nash equilibrium}
\acrodef{ofdma}  [OFDMA]  {orthogonal frequency-division multiple access}
\acrodef{cdma}   [CDMA]   {code-division multiple access}
\acrodef{ofdm}   [OFDM]   {Orthogonal frequency-division multiplexing}
\acrodef{rbs}    [RB]     {resource block(s)}
\acrodef{lte}    [LTE]    {long term evolution}
\acrodef{wimax}  [WiMAX]  {worldwide interoperability for microwave access}
\acrodef{sinr}   [SINR]   {signal to interference plus noise ratio}
\acrodef{ra}     [RA]     {resource allocation}
\acrodef{bs}     [BS]     {base station}
\acrodef{ue}     [UE]     {user equipment}
\acrodef{gt}     [GT]     {Game theory}
\acrodef{hn}     [HetNets]{heterogeneous networks}
\acrodef{sg}     [SG]     {supermodular game}
\acrodef{cdf}    [CDF]    {cumulative distribution function}
\acrodef{ue}     [UE]     {user equipment}
\acrodef{fbs}    [FBS]    {femto BS}
\acrodef{rrh}    [RRH]    {remote radio heads}
\acrodef{henb}   [HeNB]   {enhanced Home Node B}
\acrodef{fue}    [FUE]    {femto user equipment}
\acrodef{mue}    [MUE]    {macro user equipment}
\acrodef{su}     [SU]     {secondary user}
\acrodef{pu}     [PU]     {primary user}
\acrodef{pbs}    [PBS]    {primary base station}
\acrodef{sbs}    [SBS]    {secondary base station}
\acrodef{rss}    [RSS]    {received signal strength}
\acrodef{tdd}    [TDD]    {time division duplexing}
\acrodef{awgn}   [AWGN]   {additive white Gaussian noise}
\acrodef{pot}    [POT]    {Partially overlapping tones}
\acrodef{pofmt}  [POFMT]  {partially overlapping filtered multi tones}
\acrodef{fmt}    [FMT]    {Filtered multi-tones}
\acrodef{pnw}    [P'nW]   {Play\&Wait}
\acrodef{cop}    [CoP]    {continuous play}
\acrodef{rrc}    [RRC]    {root raised cosine}

\renewcommand{\figurename}{Fig.} % He DOES NOT USE IT SINCE HE USED IT MAnULAY
\renewcommand{\tablename}{Table }
\def\equname{Equ.}

\IEEEpeerreviewmaketitle

\section{Introduction}
\label{sec:intera}
\IEEEPARstart{R}{elay} selection\cite{Liu} to form cooperative communication has become a critical technology with the 5G - new radio (5G-NR) and future mobile communications. Relay selection in multi-hop communication was shown an adorable technique for mobile communication over mmWave frequency bands \cite{Qiao,Wu} The sensitivity of mmWave signal to  fading remains a fundamental challenge in communication systems especially for 5G era. Authors of \cite{Attiah} have proposed a novel adaptive multi-state selection utilizing different of mmWave frequencies. The fifth-communications generation goals are prioritized base on three pillar which are enhanced mobile broadband (eMBB), ultra-reliable, low latency communications (URLLC), and massive machine type communications (mMTC). In this work, we are trying to meet the first two goals base on relay selection with a new mechanism that increases the communication strength and improve the reliability. Moreover, this work may enhances the communication between massive devices to meet the mMTC.
During the communication propagation, the transmitted signal can be affected by the surrounding environments resulting in the signal diffraction, scattering, and reflection as showing in Figure 1.  Having a line-of-sight (LOS) transmission does not mean obtaining a proper transmission but relaying on other propagation links may have a better performance and coverage. Figure 1 demonstrates three propagated signals from (BS) where the first link $PL_{LOS}$ is the LOS signal, $PL_{Bs,r1}$ is the second transmitted signal which handover to another station and the third link is affect due to obstacles $OB$ then penetrate through to the destination mobile station $MS$. To allow the destination point to select best link base on the propagation signal strength to meet the meet the 5G -new radio (NR) requirement (the ultra-reliable and low latency communications) in term or reliability with 99.999\% \cite{142}. In this paper, we demonstrate a new machine learning methodology that acculturate the link selection from base stations or users equipment side.

\begin{figure}[!h]
	\centering
	\setlength{\abovecaptionskip}{0.1cm}
	\setlength{\belowcaptionskip}{-0.3cm}
	\includegraphics[width=0.45\textwidth]{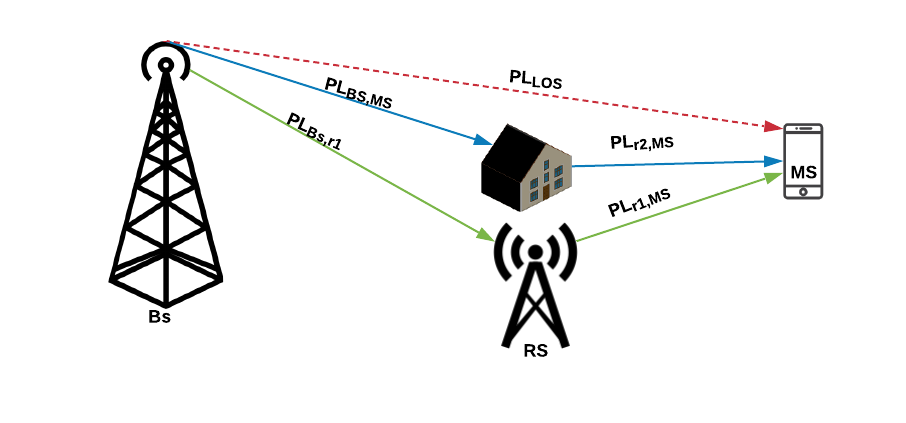}
	\renewcommand\figurename{Fig}\\
	\caption{Demonstrate the propagation links of a wireless communication where the destination user equipment (UE) selects optimum link with lower loss.}
	\label{Fig1}
\end{figure}
\begin{align}
C_i(x) = \left\{ \begin{array}{cc} 
1 & \hspace{5mm} PL<120 dBm \\
0 & \hspace{5mm} PL\geq 120 dBm \\
\end{array} \right.
\end{align}
Where $C_i(x)$ is the link selection classes which depends on the path loss of the link that can be calculated using models such as Floating-Intercept (FI) model.
\begin{equation}
PL^{FI}(f,d)[dB]=\alpha +10\beta log_{10}(d)+X^{FL}_\sigma 
\end{equation}
%\begin{equation}
%L=\argmin_{L_s}\Big((PL_{BS,r1},PL_{r1,UE}), (PL_{BS,r3},PL_{r3,UE}), .. , L_s\Big)
%\end{equation}

\begin{equation}
BL=\argmin_{L_s}(PL_{n})
\end{equation}
Where $BL$ is the best link selection and $PL_{n}$ is the path loss of the propagated links. While $n$ is the number of transmitted links between base station and destination. 

To meet the 5G pillars technologies particularity the eMBB and uRLLC, we propose a new mechanism to overcome the affected propagation link using artificial neural networks. Using relay link selection, we note that the classification of machine learning is suitable for description and prediction of the optimum link/path. Having a reliable mechanism to meet the uRLLC with trustworthy communications and eMBB to enhance the coverage and improve the communications. The supervised classification algorithms can be performed to predict categorical class labels using the training dataset of the outdoor urban environment, and can be implemented by artificial neural network (ANN) with deep learning. Furthermore, Multilayer Perceptrons (MLP) methods in ANN, together with different models, are evaluated to predict and select the strongest propagation link. MLP therefore serves the classification in ANN to identify and characterize the new link candidates using the path loss parameter or the receive signal strength. This work also will influence to the massive machine type communications (mMTC). 
 The classification technique that was selected is a binary where there are only two classes which are strong link and weak link. Once we obtain a propagation results, the new mechanism of this work divides the signal losses into categories, these levels are classes. In our case we have a binary classes, where each path loss signal strength is either considered sufficient or insufficient (no fading). The base station makes a decision based on that categorical. Thus, identifying the optimum link using a classification algorithms to meet the reliability and coverage of the 5G NR. The base station learns how to predict the weakest path loss and select the minimum loss path. The base station selects the propagated signals base on a certain energy strength (threshold) and once that link energy reaches this threshold the base station will switch to another link. In this study, the threshold is base on the path loss which is equal to -120 dBm and below this threshold is considered a poor propagation. Thus, eMBB and uRLLC will be achieved.

\cite{MaoJ2018} suggests using deep learning to identify and classify the modulation nodes, improving the interference alignment and locate the optimum routing path. Furthermore, applying prediction techniques using methods such as classification and clustering etc. to estimate the channel path loss models that lead to a better performance and precision. Multilayer Perceptrons is a ML classification technique which is a neural network. The data can be classified based on maximum probability in Multilayer Perceptrons techniques to predict the path loss that can be expressed as:
\begin{equation}
\hat{C}= \argmax_{C_i} \prod_{i=1}^{n}{P(C_i/X)}
\end{equation}
$\hat{C}$ is the prediction path loss class and $P(X_i/C)$ is the conditional probability of dataset feature given the class. \cite{Hackeling2014} published an article showing how machine learning techniques such as Deep Neural Network (DNN) reduce the complexity and increase the performance.
In this manuscript, Multilayer Perceptrons Neural Network is introduced in \rom{2}. Followed by section \rom{3} that shows  the dataset. Then, models validation and results in \rom{4}. Lastly, a conclusion is shown in section \rom{5}.

\section{Multilayer Perceptrons Neural Network}
Among many DNN structures, Multilayer Perceptrons (MLP) uses a Feed-forward neural networks (FFNNs) and a back-propagation network to compute the loss and adjust the weight \cite{Haykin}, which is suitable for deep learning. MLP forms a fully connected networks where every single node in a single layer is connected to every node in the following layers. The subsequent error is usually obtained by the loss function and optimization methods can be use to minimize the loss such as Adam optimizer. There are multiple of loss functions and cross entropy will be used when relay selection can be initially viewed as a binary classification problem. MLP is actually a multivariate multiple nonlinear regression and collection of neurons that serves as a classification by building decision decision. Multilayer Perceptrons are usually uncorrelated, and a collection of them make up the network that can be less prone to the notorious overfitting. MLP is mathematically mapping in the form of:
\begin{equation}
\Re^n \longrightarrow \Re^m: (y_1,y_2, ..., y_n)
\end{equation}
\begin{equation} 
y_n=g_{s}\Bigg(w_{0}+\sum_{i=1}^{n}w_{i}y_i\Bigg)
\end{equation}
\begin{equation} 
y_2=g_{out}\Bigg(w^{(2)}_{k0}+\sum_{j=1}^{M}w^{(2)}_{k0}\gamma\Big(w^{(1)}_{j0}+\sum_{i=1}^{n}w^{(1)}_{ji}y_i\Big)\Bigg)
\end{equation}
The above structure can proceed with only two layer, where $y_0=1$ as the output of the first layer. $g_s$ is the activation function and here we are using the step function as can be shown as:
\begin{equation}
g(\cdot):R \rightarrow R
\end{equation}
\begin{align}
g_s(x) = \left\{ \begin{array}{cc} 
0 & \hspace{5mm} x<0 \\
1 & \hspace{5mm} x\geq 0 \\
\end{array} \right.
\end{align}
To accomplish our purpose, Adam optimization algorithm will be adopted which is different than the traditional stochastic gradient descent process to update the weights iterative base in the training data \cite{Adam} with a learning rate or step size $\alpha$. Artificial intelligence (AI), particularly machine learning (ML), is widely studied to enable a system to learn of intelligence, predict and make an assessment instead of the needs of humans \cite{KC2017}. Switching the traditional link selection such as Adaptive selection scheme \cite{Shah20} to machine learning link selection still in its early stage. One of the main issues in current communications is the accuracy of handover, whereas using machine learning techniques could enhance the prediction and reduce the complexity. 
The new ML methods predict the best link using different mechanisms base on path loss and receive signal strength. \cite{Hur2001} showed how to predict the transmitted signals using deep learning techniques \cite{Hengtao} used neural networks methods such as learned decisions-based approximate message passing (LDAMP) network to estimate and learn channel state information (CSI) then solving the limited number of frequency chains in cellular systems from training data. \cite{AMolisch2018} used a model-based method using Cramer-Rao lower bound (CRLB) to predict the channel state parameters in the deeper neural network. Moreover, other authors presented in \cite{Shea2017} and \cite{Saud} some communications challenges that reach the complexity level such as atmospheric effects, handover, beam direction, MIMO and it is the time for machine learning to get evolved. 
Machine learning uses training and testing for letting the machines learn and keep predicting. In training part, learning from the data while the testing method, a trained model is used for predicting such as the ray selection. Supervised learning techniques require input, target and training data to create a model that is used for predicting. If a sample space consists of $X_i $ and output label space $y_i$ where $i ~ {1, 2,..,N}$ then by using a machine learning algorithms $\AA$, which is a function that map the input values to the labels that helps for future predicting. To measure the quality of the mapping, a loss function is used and see \cite{Mohri2012} for more details. The classification algorithms that will be used in this journey is Multilayer Perceptrons Neural Network and can be described in the following.

Multilayer Perceptrons Neural Network usually can be used to both classification and regression. When the Multilayer Perceptrons used for the classification, this algorithm works by having binary or multiple classes. However, with the regression techniques, is usually used for continuous outputs while our goal here is to classify the link strength to binary classes to predict the optimum link propagation. Multilayer Perceptrons follow the form as shown below.
\begin{equation}
y=\Phi (\sum_{i=1}^{n} w_iX_i+b)
\end{equation}
Where $w$ is the vector of weights of $x$ vector inputs and $b$ is the error. $\Phi$ is the non linear activation function. In this work, we proposed six models of Multilayer Perceptrons with different specification as shown below:
\begin{itemize}
	\item Model 1: One Hidden Layer of 10 Neurons
	\item Model 2: Two Hidden Layers of 50 and 10 Neurons
	\item Model 3: Three Hidden Layers of 10, 50 and 10 Neurons
	\item Model 4: Four Hidden Layers of 10, 50, 50 and 10 Neurons
	\item Model 5: Five Hidden Layers of 10, 50, 100, 50 and 10 Neurons
	\item Model 6: Eight Hidden Layers of 10, 50, 100, 100, 50 and 10 Neurons.
	\item Model 7: Logistic Regression Model
	\item Model 8: Dummy Classifier Model
	\item Model 9: Support Vector Machine
\end{itemize}
Multilayer Perceptrons neural network will be employed to predict the the optimum propagated link in our relay selection. Then, compare with other machine learning techniques base on precision, recall, F1 Score, accuracy and support. Results will be explored using simulated data showing the accuracy of applying deep learning learning techniques and how this algorithm performs well in relay selection. By considering more from wireless communications, 
%Couple communications issues need to be investigated with ML algorithms. Overcoming these issues improve the performance of the wireless communications. 
the result and compassion of these machine learning models to predict and selection of the best relay link will be shown later in Section IV. 

Since both prediction of link performance and classification to select an appropriate link, MLP neural networks appears fit our purpose due to capability of predicting the link with low path loss, which allows a reliable handover to meet the need of eMBB and uRLLC. While other ANN structures such as convolution neural networks are for images where there exist 2D or 3D inputs and the RNN is for sequential models like time series, machine translation, language generation. Further investigations to check the fitness of the MLP models compared to other machine learning models will be conducted later in Section IV. Thus, we proved that deep learning technique (MLP) is a capable technique to overcome this wireless communications fading issue using link selection base on MLP technique and performed better than other machine learning techniques.

\section{Dataset}
The dataset of this investigation was generated after some modification using open source Matlab simulation by New York University \cite{NYUSIM} \cite{NYUSIM2}. The dataset of the wireless channel is composed of two fragments. The selected model will be trained and validated on the dataset then will be tested using the unseen data. In this work, the train part took 75\% of the data set and 25\% for tasting the model in our training/testing scenario and others can be found in \cite{Ebhota2018}. Classes data which is zeros and ones are specified base on the path loss strength and other channel states information (CSI) are used to predict the path loss.
The measurements are specified based on distance from 1 m until 40 m. That simulation is suitable for frequencies in a range of 500 MHz to 100 GHz, bandwidth up to 800 MHz with different scenarios and environments. As a summary, simulation parameters are listed in Table \rom{1}, which exhibits that the channel measurement parameters of that data raw that was used for this paper. 
\begin{table}[!h]
	\caption{Channel Measurement Parameters.}\label{tab1}
	\begin{tabular}{| m{5 cm} | m{2cm}|} 	
		
		\hline
		
		\bfseries  Parameters &   \bfseries Values \\
		
		\hline
		Distance (m)& 1-40 \\
		
		\hline
		Frequency (GHz) & 28 \\
		
		\hline
		Bandwidth (MHz) & 800\\
		
		\hline
		TXPower (dBm) & 30\\
		
		\hline
		Scenario & UMi \\
		
		\hline
		Polarization & Co-Pol \\
		
		\hline
		TxArrayType & ULA \\
		
		\hline
		RxArrayType & ULA \\
		
		\hline
		Antena & SISO \\
		
		\hline
		Tx/Rx antenna Azimuth and Elevation (red) & 10\\
		
		\hline
	\end{tabular}
\end{table}

The dataset that used in this work are consisting of channel properties of a communications link such that the information helps the base station to execute supervised classification based on a dataset from prior measurements or simulations.

\section{Models Validation and Results}
To accomplish a broad exploration, 
Multilayer Perceptrons Neural Network, Logistic Regression, Dummy Classifier and Support Vector Machine are used to perform the classification techniques. Evaluating the performance of these classification algorithms by confusion matrix which counts the outcomes of the prediction models compared to training dataset \cite{Rud}. Moreover, the precision usually shows how often a model make a positive prediction and recall shows how the model is confident of the predicting all positive targets. Accuracy, Precision, Recall, and F1 Score metrics were used to evaluate the machine learning classification algorithms (classifiers). The accuracy is measured by counting the number of true prediction to the total number of predictions. Which is the number of correctly predicted selected links over the total number of links, which tells how the classifier is able not to misclassify a positive path loss (a sample). Precision is the number of true positives ($T_p$) over the number of true and false positives ($F_p$). Recall stands for the number of true positives over the number of true positives and false negatives (FN). While F1 Score measures the harmonic mean for both precision and recall, we obtain the following mathematical expressions:
\begin{equation}
\mbox{Average Precision} = \frac{1}{n}\sum_{i=1}^{N} \frac{Tp}{Tp+Fp}
\end{equation}
 \begin{equation}
\mbox{Total Recal} = \sum_{i=1}^{N} \frac{Tp}{Tp+FN}
 \end{equation}
\begin{equation}
F_1 \mbox{Score} =  2 \times \frac{precision \times recall}{precision + recall}
\end{equation}
\begin{table}[!h]
	
	\caption{Interpretation of Performance Measures.}\label{tab1}
	\begin{tabular}{| m{3.1 cm} | m{1.3cm}|  m{1.3cm}|  m{1.3cm}|} 
		
		\hline
		\bfseries  ANN Models&   \bfseries Precision  &   \bfseries Recall &   \bfseries F1 Score \\
		
		\hline
		\textbf{Model 1} & 0.39   &   0.61  &    0.47\\
		
		\hline
		\textbf{Model 2} &  0.88   &   0.87  &    0.87  \\
		
		\hline
		\textbf{Model 3 }&  0.86  &    0.86    &  0.86 \\
		
				\hline
		\textbf{Model 4 }& 0.93    &  0.91   &   0.92 \\
		
				\hline
		\textbf{Model 5 }& 0.98   &   0.98    &  0.98  \\
					\hline
		\textbf{Model 6 }& 0.88   &   0.87   &   0.88  \\
		
							\hline
		\textbf{Logistic Regression  }& 0.86   &   0.86   &   0.86  \\
		
							\hline
		\textbf{Dummy Classifier }& 0.56   &   0.57   &   0.57  \\
		
							\hline
		\textbf{SVM }& 0.92  &   0.93    &  0.93   \\
		\hline
	\end{tabular}
\end{table}
\begin{table}[!h]
	\caption{Accuracy compression of all models.}\label{tab1}
	\begin{tabular}{| m{3 cm} | m{1.5cm}|  m{2.90cm}|} 	
		
		\hline
		\bfseries   Models &     \bfseries Accuracy &   \bfseries ROC AUC Score\\
		
		\hline
		\textbf{Model 1} & 0.623 &	0.484 \\
		
		\hline
		\textbf{Model 2} & 	0.868 &	0.877 \\
		
		\hline
		\textbf{Model 3}& 	0.857 &	0.842\\
		
			\hline
		\textbf{Model 4 }& 	0.925 &	0.932\\
			\hline
		\textbf{Model 5 }& 	0.982 &	0.981\\
			\hline
		\textbf{Model 6 }& 	0.882 &	0.866\\
		
			\hline
		\textbf{Logistic Regression }& 	0.882 &	0.866\\
		
			\hline
		\textbf{Dummy Classifier }& 	0.857 &	0.848\\
		
			\hline
		\textbf{SVM }& 	0.934 &	0.973\\

		\hline
	\end{tabular}
\end{table}

The interpretations of performance measures that were used to check the process of a model via precision, recall, f1 Score, accuracy and support. From the two above tables interpretation of performance measures of the Multilayer Perceptrons Neural Network algorithms, we can conclude that all of these techniques did a decent job and model 5 has the highest accuracy which consist of five Hidden Layers of 10, 50, 100, 50 and 10 Neurons.
Model 5 gained in the best performance in Precision, Recall, and F1 Score  among other models. Thus, it has the best performance in classifying the the ray selection, followed by the Model 4 and and worst one is model 1 among the ANN model and the dummy classifier compared to all models. The reason for that is some of the features depends on each other such as distance and received power. Electing the number of hidden layers and number of neurons are still an open research topic where few or more neurons leads to underfitting and overfitting. An assumption from our trail and error trails, we noticed that the number of hidden layers should lower than the number of input by 30\%. Model 6, began degrading once the number of hidden layers have reached 70\% of the number of the inputs as can be seen in figure 2. 
\begin{figure}[!h]
	\centering
	\setlength{\abovecaptionskip}{0.1cm}
	\setlength{\belowcaptionskip}{-0.3cm}
	\includegraphics[width=0.45\textwidth]{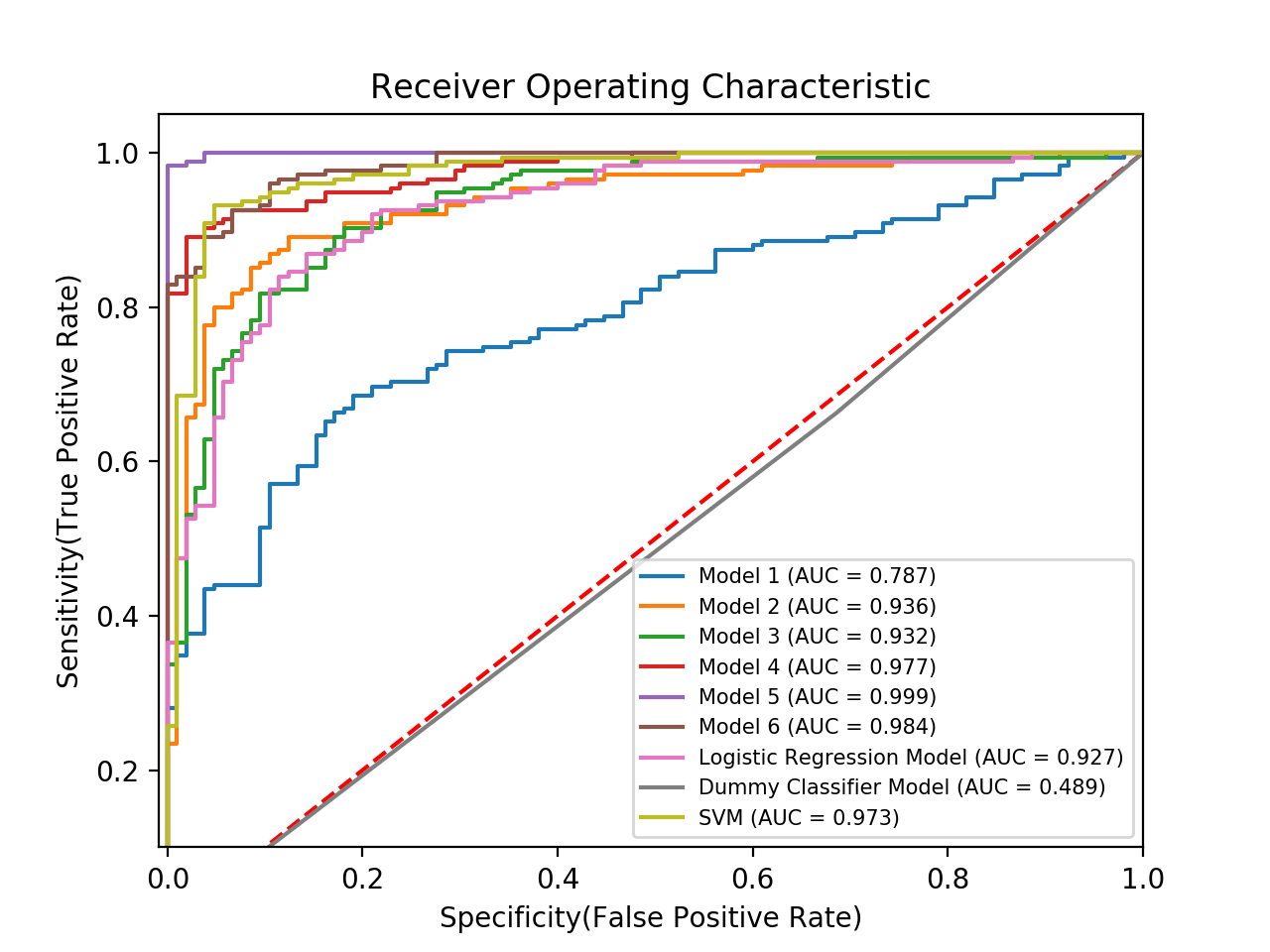}
	\renewcommand\figurename{Fig}\\
	\caption{ROC curves of Classification techniques.}
	\label{Fig3}
\end{figure}
\begin{figure}[!h]
	\centering
	\setlength{\abovecaptionskip}{0.1cm}
	\setlength{\belowcaptionskip}{-0.3cm}
	\includegraphics[width=0.45\textwidth]{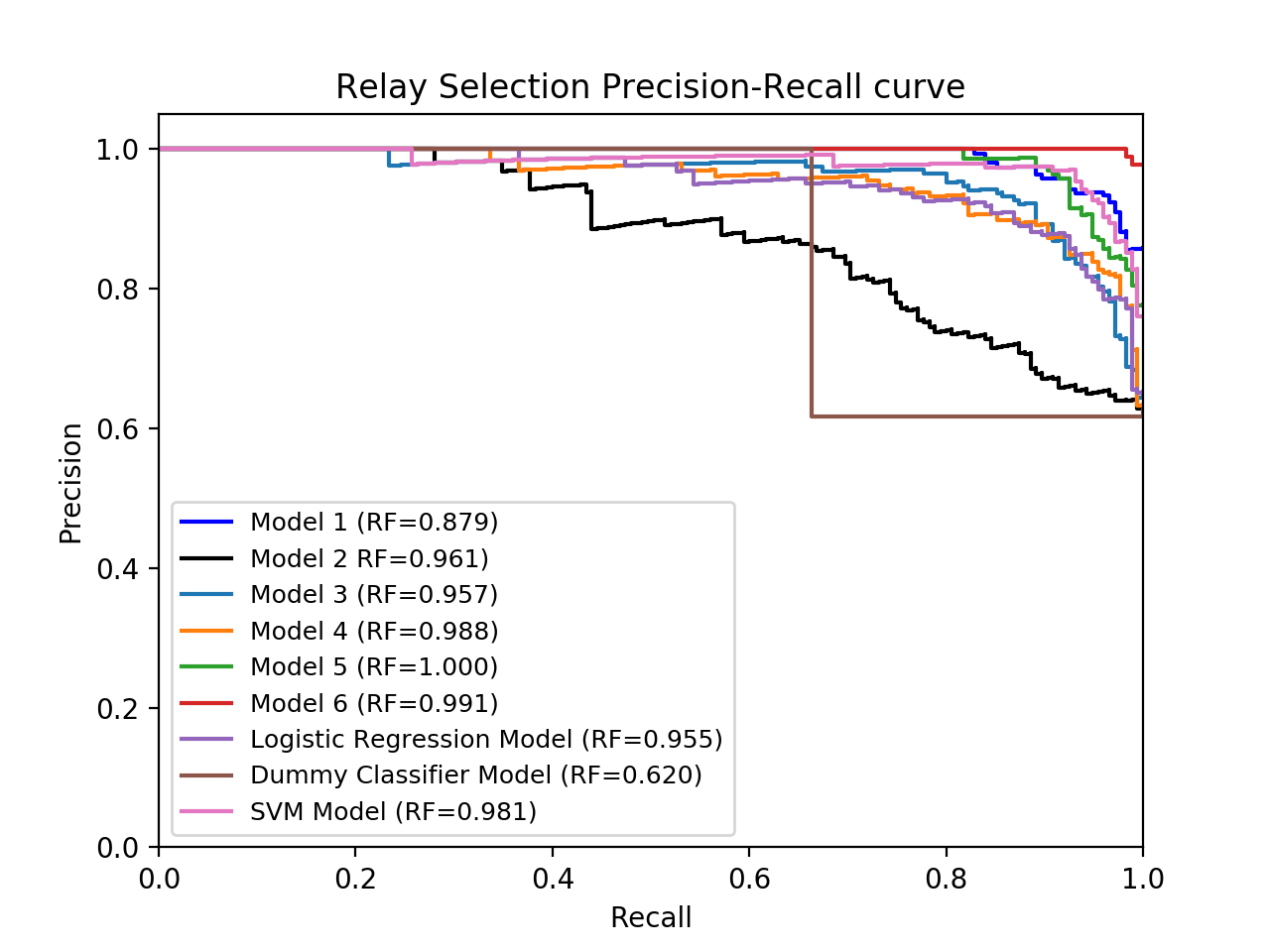}
	\renewcommand\figurename{Fig}\\
	\caption{Precision and recall curve  of Classification techniques.}
	\label{Fig3}
\end{figure}
\begin{figure}[!h]
	\centering
	\setlength{\abovecaptionskip}{0.1cm}
	\setlength{\belowcaptionskip}{-0.3cm}
	\includegraphics[width=0.45\textwidth]{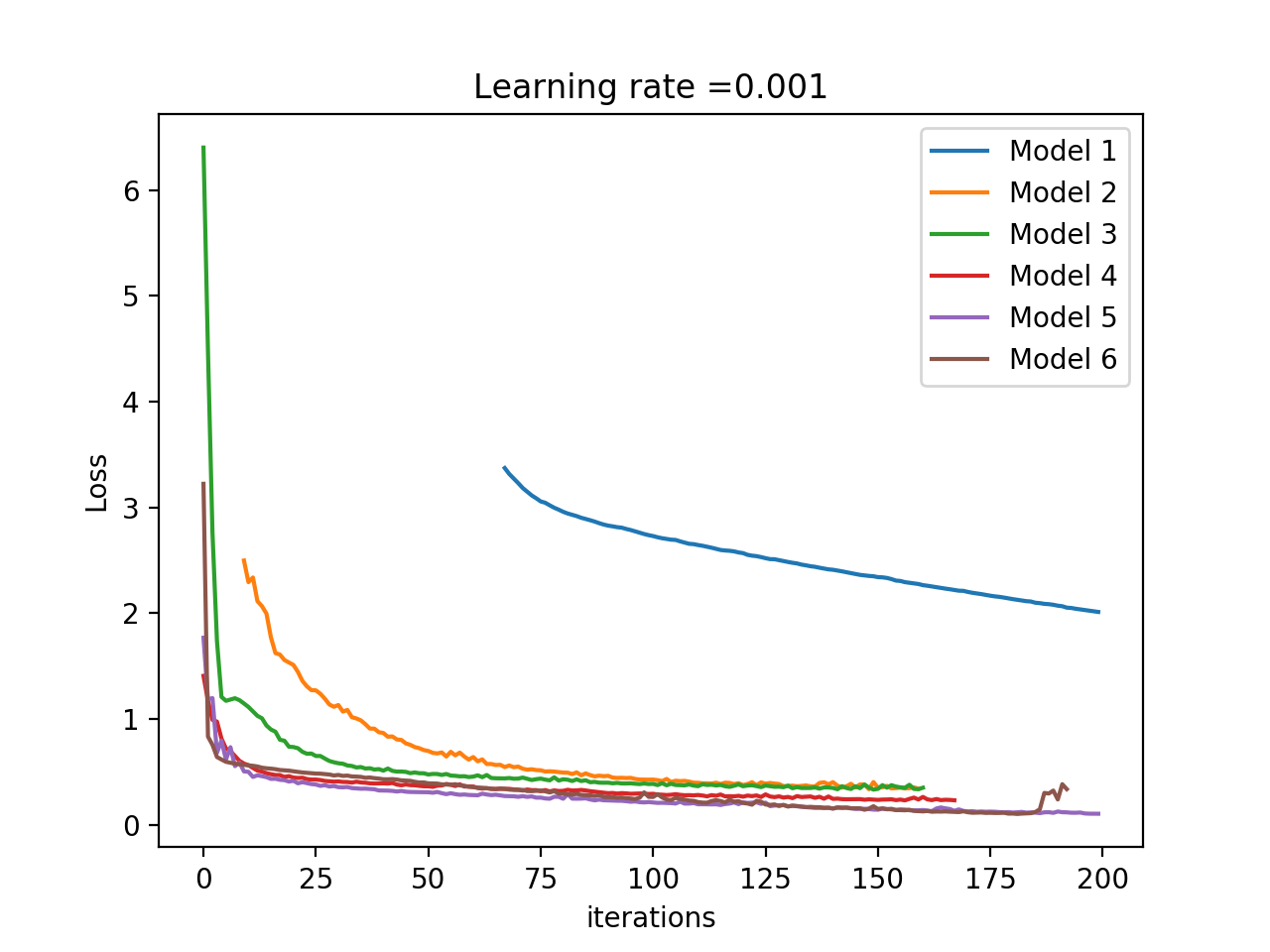}
	\renewcommand\figurename{Fig}\\
	\caption{Loss VS iteration of ANN Classification techniques.}
	\label{Fig3}
\end{figure}
Figure 2 presents the Receiver Operating Characteristic (ROC) \cite{Powers} visualizes a classifier's performance that illustrates the wellness of the classification models used in this paper. The false positive rate is plotted against the true positive rate and it's very obvious that the model 5 is almost closer to the optimum with 99\%. Other models have been compared our models such support vector machine (SVM) that was presented in \cite{SVM} that could not preform well in overcoming the relay link selection. The usage of sigmoid kernel while using the linear kernel poly kernel with degree 4 would improve the performance. Underfitting and overfitting were examined base on checking the accuracy and the ROC score of the training and the testing data which were showing close to each other and performing good.
Figure 3 illustrates the relay selection precision recall curve where model 5 is performing the best among other models. The precision recall curve shows the relationship between true positive rate and the positive prediction value for a varieties of models. While figure 4 shows the history loss verse neural iterations when the training data will not improve the performance of the model by at least tolerance (usually assigned 1e-4) or having a constant loss for multiple of iteration. From the figure, we notice the losses of models decreased nicely and smoothly except model one due the adjusted learning rate of this model which was 1e-5 while others are 0.05 and increase that rate will affect the accuracy of the model. Moreover, by looking at model 6, we notice at iteration number 185, it starts increasing which is a sign to stop the model to avoid issues such as overfitting and decreasing the efficiency of the model. Figure 4 again confirm the best performance goes to model 5 among other NN models.
Simply we conclude this journey from the result section that ANN performs better in term of selecting the optimum link comparing to other machine learning techniques where the accuracy of selecting the optimum link is 99\% which meet third 5G -new radio (NR) requirement (the ultra-reliable and low latency communications).
The future work to develop are extraordinarily rich, and powerfully using machine learning algorithms toward solving wireless communications issues to enhance the the communications and reduce the complexities for future wireless generations. This approach can be further extended to solve other wireless communications issues such as near-far problem base on more than a binary classification. Base on a certain path loss, power transmission control can be manipulated to achieve a full communications efficiency.
\section{Conclusion}
Wireless communications with the new era of 5G and beyond require overcome classical issues in order to meet the 5G pillars that includes eMBB, uRLLC and mMTC. One of these critical issues is relay selection to handover with reliability to the strongest link to meet the eMBB and uRLLC. This can be solved by applying machine learning techniques such as Multilayer Perceptrons Neural Network, Logistic regression, Dummy classifier and support vector machine to develop alternative techniques to predict the signal path loss strength that's usually get affected by the wireless channel. Multilayer Perceptrons is classification technique used in this journey and compared the result of each model by using the interpretation of performance measures such as accuracy, precision, recall and F1-score and compare with previous studies to end up with a better performance. Other techniques influenced a perfect predication that confirm the usage of machine learning towards wireless communications.
\section{Acknowledgement}
Saud Aldossari expresses a great appreciation to Prince Sattam bin Abdulaziz University for their support of providing scholarship. K.-C. Chen appreciates the support from Cyber Florida. 
%\vspace {10 pt}

\balance  %% to go to next column only at end needed for the referance sake of arrangments

\bibliographystyle{IEEEtran}
%\bibliography{saud.bib}

\end{document}